\documentstyle[12pt]{article}
\renewcommand{\theequation}{\arabic{equation}}
\renewcommand{\thesection}{\arabic{section}}
\renewcommand{\thefootnote}{\fnsymbol{footnote}}
\newcommand{\bea}{\begin{eqnarray}}
\newcommand{\ena}{\end{eqnarray}}
\newcommand{\vs}[1]{\vspace{#1 mm}}

\renewcommand{\a}{\alpha}
\renewcommand{\b}{\beta}
\renewcommand{\c}{\gamma}

\newcommand{\e}{\epsilon}
\newcommand{\n}{\nu}
\newcommand{\z}{\omega}
\newcommand{\k}{\kappa}
\renewcommand{\t}{\tau}
\newcommand{\G}{\Gamma}

\newcommand{\CMP}[1]{Comm.\ Math.\ Phys.\ {\bf #1}}
\newcommand{\PR}[1]{Phys.\ Rev.\ {\bf #1}}
\newcommand{\PRL}[1]{Phys.\ Rev.\ Lett.\ {\bf #1}}
\newcommand{\PTP}[1]{Prog.\ Theor.\ Phys.\ {\bf #1}}
\newcommand{\AJ}[1]{Astorophys. \ J.\ {\bf #1}}
\newcommand{\JMP}[1]{J.\ Math.\ Phys.\ {\bf #1}}

\begin{document}
\noindent
\topmargin 0pt
\oddsidemargin 5mm

\begin{titlepage}
\setcounter{page}{0}
\begin{flushright}
November, 1996\\
OU-HET 254\\
gr-qc/9611014\\
\end{flushright}
\vs{10}
\begin{center}
{\Large{\bf Analytic Solutions of the Teukolsky Equation and their 
Properties}}\\
\vs{10}
{\large  
Shuhei Mano,\footnote{e-mail address: mano@phys.wani.osaka-u.ac.jp}
and Eiichi Takasugi\footnote{e-mail address: 
takasugi@phys.wani.osaka-u.ac.jp}
}\\
\vs{8}
{\em Department of Physics,
Osaka University \\ Toyonaka, Osaka 560, Japan} \\
\end{center}
\vs{10}
\centerline{{\bf Abstract}}  
The analytical solutions reported in our previous paper are given as 
series of hypergeometric or Coulomb wave functions. By using them, 
we can get the Teukolsky functions analytically in a desired accuracy. 
For the computation,  the deep understanding  of their  properties is 
necessary.  We summarize the main result: The relative normalization 
between the solutions with a spin weight $s$ and $-s$ is given 
analytically by using the Teukolsky-Starobinsky (T-S) identities. 
By examining the asymptotic behaviors of our solution and 
combined with the T-S identities and the Wronskian, we found 
nontrivial identities between the sums of coefficients of the series. 
These identities will serve to make various expression in simpler 
forms and also become a powerful tool to test the accuracy of the 
computation. As an application, we investigated the absorption rate and 
the evaporation rate of black hole and obtain interesting analytic results.

\end{titlepage}

\newpage
\renewcommand{\thefootnote}{\arabic{footnote}}
\setcounter{footnote}{0}

\section{Introduction}
In our previous paper[1], we reported the analytic solutions of Teukolsky 
equation[2] which consist of two type of series; one is given 
in the form of series of hypergeometric functions (hereafter we 
call it the hypergeometric type solution) and the other is given in the 
form of series of Coulomb wave functions (hereafter we call it the Coulomb 
type solution which was first given by Leaver[3]). 
The hypergeometric type solutions are shown to be 
convergent in the region except infinity with all finite $\e=2M\omega$, $M$ 
being the mass of black hole and $\omega$ an angular frequency. The 
Coulomb type solutions are convergent in the region far above the outer 
horizon  for all finite $\e$. 
We showed that the  matching of these two types of solutions is perfect 
in the intermediate region where both solutions are convergent. In this way, 
we found the solutions which are convergent in the entire region of $r$ 
for all finite $\e$. In addition, we showed that our solutions are 
suitable to get the Teukolsky functions in the $\e$ expansion or the 
numerical computation. We also presented analytical solutions of 
Regge-Wheeler equation[4].

We showed[1] that the solution can be written in  various forms so that 
we have to choose some special ones. Furthermore, it is necessary to 
investigate further the properties of solutions to calculate the 
Teukolsky functions, because the solutions are given as series. 
For this purpose, we examined the asymptotic behaviors of the solutions 
and compared with the general result derived before. We made a systematic 
study on  the incoming solution on the outer horizon.

As for the incoming solution on the horizon, we take 
the same hypergeometric type one as in Ref.1, for which 
the boundary condition on the horizon is trivial. 
For the outgoing solution, we take a different form. 
As for the  Coulomb type solutions, we take different forms 
from that given in Ref.1. In this paper, we choose a variable 
$z=\omega (r-r_+)$ which is different from the definition in Ref.1 
where we took$z=\omega (r-r_-)$.  Accordingly, the solutions adopted 
here are different. We took them  because  
$z$ reduces to $\omega r$ in the Schwarzschild limit and thus 
the solutions for large $r$ can be compared  with  
the previous analysis by Tagoshi and Nakamura, and  Sasaki and companies[5] 
for Schwarzshild case and to our solutions of the Regge-Wheeler equation[4]. 
Therefore, some of solutions given here are new ones. 
 
In order to investigate the properties of the solutions, we mainly 
examine the Teukolsky-Starobinsky (T-S) identities[6],[7] 
and the Wronskian[6]. 
Although the solutions of Teukolsky equation satisfies the 
T-S identities automatically, it is non trivial 
to see directly that our solutions satisfy them because our solutions 
are given as series. Fortunately, we were able to show analytically that 
the incoming solution satisfies one of T-S identities. 
By using this, we can fix analytically 
the relative normalization between solutions with a spin weight $s$ and $-s$. 
It was hard to prove that our solution satisfies another T-S 
identities directly. In order to investigate the T-S identities and 
the Wronskian,  we considered  the asymptotic behaviors. For this, we 
derived the expressions of asymptotic amplitudes on the horizon  and 
at infinity from our  solution. By examining  the T-S identities and 
the Wronskian, we found two very nontrivial identities among sums of 
coefficients.  Since the solutions are defined as series, various 
quantities expressed by Teukolsky functions necessary involve the sums 
of coefficients which are calculated by solving the three term 
recurrence relation. These identities will serve to make  the 
expressions of them in compact forms form which we may be able to 
have a direct incite of the physical meaning. As an example, we 
considered the absorption rate and the evaporation rate of black hole 
and obtained compact forms of them. 
For computing Teukolsky functions, we will make either  the 
 $\e$ expansion or the numerical computation of coefficients. 
In order to test the accuracy of the computation, we can use these 
identities. Since the sum of coefficients in these identities are 
proportional to the asymptotic amplitudes on horizon or at infinity, 
the check of these identities gives a direct check of accuracy of 
Teukolsky functions.  

With these understandings of the solutions given in this paper, 
it is  now in the position to  make  the numerical study of 
Teukolsky functions. By solving the three term recurrence relation, 
we can in principle obtain Teukolsky functions with a desired accuracy. 
The further numerical study is in pressing need.

In Sec.2, we give the outline of solutions of Teukolsky equation 
and new forms of them. In Sec.3, various 
properties of coefficients the proportionality factor between 
the hypergeometric type and Coulomb type solutions are given. 
 The asymptotic behaviors 
on the outer horizon and at infinity are expressed directly by using 
our solution in Sec.4. In Sec.5, we discuss properties of our solution  
by using  the T-S identities. By using the Wronskian, we discuss  
the conserved quantity and the energy conservation in Sec.6. By using 
the T-S identities and the Wronskian, we derive two identities among 
the sums of coefficients in Sec.7.  The absorption rate and the evaporation 
rate by black hole are discussed in Sec.8.  
Discussions are given in Sec.9. The outlines of proofs of various formulas 
in the text are given in Appendix A.

\section{Analytic solutions} 
 
The hypergeometric type solution which we discuss here is the same as 
the one given in Ref.1. However, we take different solutions for the Coulomb 
type from the one in Ref.1. We do not present the derivations and the proofs 
of various equations and statements which one can find  in Ref.1 or 
one can  prove following the discussions given there. 

\vskip 3mm
\noindent
(a) Notation 

We start from the Teukolsky function[2],[6]
\bea
\Upsilon_s \simeq e^{-i\omega t} e^{im \phi}{}_sS_{l}^m (\theta)
{}_sR_{\z l m}(r),
\ena 
where 
${}_sS_{l}^m (\theta)$ is a spin-weighted spheroidal function 
with a spin-weight parameter $s$ and $l$ is an angular momentum 
which satisfies $l \ge \max(\mid m \mid, \mid s \mid)$, and 
 ${}_sR_{\z l m}(r)$ is a radial function which we write  $R_s$ for 
simplicity. 
Solutions of the radial equation are expressed as functions of 
variables $x$ and $z$ which are defined by  
\bea
 x=-\frac{\omega}{\e\k} (r-r_+), \hskip 5mm {z}=\omega (r-r_-)
=\e\k (1-x)=\e \kappa \tilde{x},
\ena
where 
\bea
\e=2M\omega,  \hskip 5mm \kappa={\sqrt{1-q^2}}, \hskip 5mmq={a\over{M}}\;.
\ena 
Here, $M$ and $a$ are the mass and the angular momentum of 
the Kerr black hole, respectively.   The parameters 
$r_{\pm}$ are poles of $\Delta=r^2+a^2-2Mr$ and is expressed by 
$r_{\pm}=M\pm\sqrt{m^2-a^2}=\e(1\pm \k)/2\omega$ where 
$r_{\pm}$ give positions of the outer and the inner horizon. 
For later use, it is convenient to define  
\bea
\t={{\e-mq}\over{\kappa}}, \hskip 5mm k=\omega-\frac{ma}{2Mr_+}=
2\frac{\k}{1+\k}\left( \frac{\omega}{\e}\right )\e_+
\ena
and 
\bea
\e_{\pm}=\frac{(\e\pm\t)}{2}\;  . 
\ena

\vskip 3mm
\noindent 
(b) Coefficients 

Solutions of the radial part are expressed in the form of series of 
hypergeometric 
functions and also in the form of series of Coulomb wave functions. 
Coefficients of these series  satisfy the same three term recurrence 
relation[1],[8] 

\bea
\a_n^{\n}(s) a_{n+1}^{\n}(s)+\b_n^{\n}(s) a_n^{\n}(s)
+\c_n^{\n}(s) a_{n-1}^{\n}(s)=0,
\ena
where
\bea
\a_n^{\n}(s)&=&{i\e \kappa (n+\n+1+s+i\e)(n+\n+1+s-i\e)(n+\n+1+i\t)
\over{(n+\n+1)(2n+2\n+3)}},
\nonumber\\
\b_n^{\n}(s)&=&-\lambda_s-s(s+1)+(n+\n)(n+\n+1)+\e^2+\e(\e-mq)\nonumber\\
&&\hskip 3mm +{\e (\e-mq)(s^2+\e^2) \over{(n+\n)(n+\n+1)}},
\nonumber\\
\c_n^{\n}(s)&=&-{i\e \kappa (n+\n-s+i\e)(n+\n-s-i\e)(n+\n-i\t)
\over{(n+\n)(2n+2\n-1)}},
\ena 
where 
$\lambda_s=E(s)-2am\omega+a^2\omega^2-s(s+1)$ with $E(s)$ being the eigenvalue 
of the spin-weighted function ${}_sS_{l}^m (\theta)$.

The series of coefficients converges if the transcendental equation 
for $\n$ 
\bea
R_n(\n)L_{n-1}(\n)=1,  
\ena
is satisfied, where $R_n(\n)$ and $L_{n}(\n)$ are the continued 
fractions defined by 
\bea
R_n (\n)&=&{a_n^{\n}(s)\over{a_{n-1}^{\n}(s)}}=
-{\c_n^{\n}(s)\over{\b_n^{\n}(s)+\a_n^{\n}(s)R_{n+1}(\n)}},
\nonumber\\
L_n(\n)&=&{a_n^{\n}(s)\over{a_{n+1}^{\n}(s)}}=
-{\a_n^{\n(s)}\over{\b_n^{\n}(s)+\c_n^{\n}(s)L_{n-1}(\n)}}\;.
\ena
The solutions are characterized $\n$ which we called  the renormalized 
(shifted) angular momentum  because $\n=l+O(\e^2)$ as shown in Ref.1. 
We can prove that if $\n$ satisfies Eq.(8), then $-\n-1$ also satisfies, 
i.e., 
$ R_n(-\n-1)L_{n-1}(-\n-1)=1$. In other words, if $\n$ is a 
solution, then $-\n-1$ is also a solution. Note that $E(s)$ is an 
even function of $s$, $E(-s)=E(s)$ as shown by Press and Teukolsky[9]. 
By using this, we can  prove that $\n$ is an even function of $s$
\bea
\n (s)=\n (-s),
\ena
which is important and enables us to relate the solution of spin weight 
$s$  to 
that $-s$   by the Teukolsky-Starobinsky identities. 
This can be proved by noticing that the transcendental equation (8) 
contains $\b_n^{\n}(s)$ and a combination $\a_k^{\n}(s)\c_{k+1}^{\n}(s)$ 
which are even functions of $s$. Thus, the equation (8) is invariant 
under the change of $s$ to $-s$ and  its eigenvalue $\n(s)$ is 
an even function of $s$. 

For $\n$ which satisfies  Eq.(10),  we find  
\bea
\displaystyle \lim_{n \rightarrow \infty} n\frac{a_n^{\n}(s)}
{a_{n-1}^{\n}(s)}=
-\lim_{n \rightarrow -\infty}  n\frac{a_n^{\n}(s)}{a_{n+1}^{\n}(s)}
=\frac {i\e \kappa }2,
\ena
which enables to establish the region of convergence of solutions. 

\vskip 3mm
\noindent
(c) Analytic solutions

\vskip 3mm
\noindent
(c-1) Hypergeometric type solutions
  
The incoming solution is given by 
\bea
R_{{\rm in};s}^{\n}&=&A_s  e^{i\e\kappa x}(-x)^{-s-i\e_+}(1-x)^{i\e_-}
\sum_{n=-\infty}^{\infty}a_{n}^{\n}(s)\nonumber\\
&&\hskip 5mm \times F(n+\nu+1-i\t,-n-\n-i\t;1-s-2i\e_+;x)\;,
\ena
where $a_{n}^{\n}(s)$'s are coefficients which satisfy the three 
term recurrence relation in Eq.(6) and 
$A_{ s}$ is the normalization  constant which we take for 
$s>0$  with the choice $A_{-s}=1$
\bea
A_{-s}=1, \hskip 5mm A_{s}=C_{s}^*
\left ( \frac{\omega}{\e \kappa} \right )^{2s}
 \frac{\G(1+s-2i\e_+)}{\G(1-s-2i\e_+)}
\left| \frac{\G(\n+1-s+i\e)}{\G(\n+1+s+i\e)}\right|^2\;,
\ena 
where $C_s$'s are Starobinsky constants (defined in Eqs.(49)-(52)). 
The relative normalization is determined   
so that the Teukolsky-Starobinsky identities are satisfied as 
we shall see later. 
It may be worthwhile to mention that 
$R_{{\rm in};s}^{-\n-1}=
R_{{\rm in};s}^{\n}$ so that the change $\n$ into $-\n-1$ does not 
lead to  a new solution. 

Another independent solution is the  
outgoing solution on the outer horizon and  is given by 
\bea 
R_{{\rm out};s}^{\n}= \Delta^{-s}(R_{{\rm in};-s}^{\n})^*\;
\ena 
which is  different  from the one given  in Ref.1. 
Solutions, $R_{{\rm in};s}^{\n}$ and $R_{{\rm out};s}^{\n}$ 
form a pair of independent solutions of Teukolsky equation. 

By  using large  $|n|$  behaviors of coefficients in Eq.(11) 
and hypergeometric functions, we can prove (see Ref.1)  
 that  the solution in Eq.(12)  is  
convergent for the entire complex plain except infinity for all 
finite $\e$. 
If we restrict $x$ to the physical region, the convergence region is 
$ 0 \le (-x) < \infty $, i.e., $r<\infty$. 

The solution can be analytically continued by  
\bea
R_{{\rm in};s}^{\n}&=&
A_s 
e^{i\e\kappa}\left[ { R}_{0,s}^{\n}+{ R}_{0,s}^{-\n-1}\right ]\;,
\ena
where 
\bea
{ R}_{0;s}^{\n} 
&=&e^{-i\e\kappa \tilde{x}}(\tilde{x})^{\n+i\e_+}
(\tilde{x}-1)^{-s-i\e_+}\nonumber\\
&&\times \sum_{n=-\infty}^{\infty}\frac{\G(1-s-2i\e_+)\G(2n+2\n+1)}
{\G(n+\n+1-i\t)\G(n+\n+1-s-i\e)}a_{n}^{\n}(s)
\nonumber\\
&&  \times 
\tilde{x}^n F(-n-\n-i\t,-n-\n-s-i\e;-2n-2\n;\frac 1{\tilde{x}}) \;.
\nonumber\\
\ena
Then, ${ R}_{0;s}^{\n}$ and  ${ R}_{0;s}^{-\n-1}$ 
form a pair of  independent solutions. 

\vskip 3mm
\noindent
(c-2) Coulomb type solutions 

The solution is given by
\bea
{R}_{C;s}^{\n}&=& 
{z}^{-1-s}\left(1-{\e \kappa \over{{z}}}\right)^{-s-i\e_+}
 \sum_{n=-\infty}^{\infty}(-i)^n\frac{(\n+1+s-i\e)_n}{(\n+1-s+i\e)_n}
a_n^{\n}(s)F_{n+\n,s}({z})\;,
\nonumber\\
\ena
where $(a)_n=\G(n+a)/\G(a)$ and 
\bea
F_{n+\n,s}(z)&=&e^{-iz}(2z)^{n+\n}z
{\Gamma(n+\n+1-s+i\e)\over{\Gamma(2n+2\n+2)}}\nonumber\\
&&\hskip 1cm \times 
\Phi(n+\n+1-s+i\e,2n+2\n+2;2iz)\;.
\ena 
This solution and the definition of $z$ are  different from 
those given in Ref.1.  
It is amusing to observe that coefficients appearing in the hypergeometric 
type solution are the same as those in the Coulomb type solutions[1],[9]. 
Thus, the renormalized (shifted) angular momentum takes the same value both 
for the hypergeometric type and the Coulomb type solutions. 
This fact is quite important for matching of these two different types of 
solutions  in the  region where both 
solutions are  convergent. 

Another independent solution is obtained by changing $\n$ to $-\n-1$,
\bea
R_{C;s}^{-\n-1}\;.
\ena
Solutions ${R}_{C;s}^{\n}$ and ${R}_{C;s}^{-\n-1}$ form  
a pair of independent solutions of Teukolsky equation. 

Coulomb type solutions given above contain both the incoming and the 
outgoing solutions at infinity. Thus, solutions are decomposed into another 
pair of solutions, the incoming solution at infinity $R_{+;s}^{\n}$   
and the outgoing solution at infinity $R_{-;s}^{\n}$. 
Explicitly, we have 
\bea
R_{C;s}^{\n}=R_{+;s}^{\n}+R_{-;s}^{\n},
\ena
where 
\bea
R_{+;s}^{\n}&=& 2^{\n}e^{-\pi \e}e^{i\pi(\n+1-s)}
\frac{\G(\n+1-s+i\e)}{\G(\n+1+s-i\e)}
e^{-iz}z^{\n+i\e_+}(z-\e\k)^{-s-i\e_+}\nonumber\\
&\times&  \sum_{n=-\infty}^{\infty}i^n
a_n^{\n}(s)(2z)^n\Psi(n+\n+1-s+i\e,2n+2\n+2;2iz),\nonumber\\
&&\\
R_{-;s}^{\n}&=& 2^{\n}e^{-\pi \e}e^{-i\pi(\n+1+s)}
e^{iz}z^{\n+i\e_+}(z-\e\k)^{-s-i\e_+}\sum_{n=-\infty}^{\infty}i^n\nonumber\\
&\times&  \frac{(\n+1+s-i\e)_n}{(\n+1-s+i\e)_n}
a_n^{\n}(s)(2z)^n\Psi(n+\n+1+s-i\e,2n+2\n+2;-2iz)\;.\nonumber\\
 \ena

By using 
the large $\mid n \mid$ behaviors of coefficients in Eq.(11) and Coulomb 
wave functions, we can prove (see Ref.1) that solutions are 
  convergent in the complex region of  
 $ \mid z\mid > \e\k$ for all finite $\e$. 
If we confine to the physical region, the 
convergence region is $r > r_+$. 

\vskip 3mm
\noindent
(c-3) Matching of two types of solutions

Let us  compare ${ R}_{0;s}^{\n}$  and ${R}_{C;s}^{\n}$. We first 
observe that they 
are solutions of Teukolsky equation and behaves  like  
$\tilde x^{\n}$ multiplied by a single-valued function 
for large $\tilde x$ ($z=\e\k \tilde x$). Thus, 
${ R}_{0;s}^{\n}$  should be proportional to  ${R}_{C;s}^{\n}$.
The same is true for  
${ R}_{0;s}^{-\n-1}$  and ${R}_{C;s}^{-\n-1}$. 
It should be noted that $\n -(-\n-1)=2\n+1$ is not an integer so that 
${R}_{0;s}^{\n}$ and ${R}_{0;s}^{-\n-1}$ are solutions with different 
characteristic exponents. 

We define 
\bea
{ R}_{0;s}^{\n} =K_{\n}(s){R}_{C;s}^{\n}\;.
\ena 
Then,  by comparing each power of  $\tilde x$  in the region 
$1<< \tilde x < \infty$ where both solutions converge,    
 we find  
\bea
{K}_{\n}(s)&=&
\frac{(2\e \kappa )^{-\n+s-{\tilde r}}2^{-s}i^{\tilde r} 
\G(1-s-2i\e_+)\G({\tilde r}+2\n+1)\G({\tilde r}+2\n+2)}{
\G(\n+1-i\t)\G(\n+1-s-i\e)\G({\tilde r}+\n+1-s+i\e)} \nonumber\\
&&\times 
\frac{\G(\n+1+i\t)\G(\n+1+s+i\e)}{
\G({\tilde r}+\n+1+i\t)\G({\tilde r}+\n+1+s+i\e)}\nonumber\\
&&\hskip 5mm \times \left ( \sum_{n={\tilde r}}^{\infty}
\frac{({\tilde r}+2\n+1)_n}{(n-{\tilde r})!}
\frac{(\n+1+s-i\e)_n}{(\n+1-s+i\e)_n}a_n^{\n}(s) \right )^*
\nonumber\\
&&\hskip 5mm \times \left(\sum_{n=-\infty}^{{\tilde r}}
\frac{(-1)^n}{({\tilde r}-n)!
({\tilde r}+2\n+2)_n}\frac{(\n+1+s-i\e)_n}{(\n+1-s+i\e)_n}
a_n^{\n}(s)\right)^{-1}\;,\nonumber\\
\ena
where $\tilde r$ can be any  integer  and 
${K}_{\n}(s)$ is independent of the choice of  ${\tilde r}$. 

\vskip 3mm
\noindent
(c-4) Solutions valid in the entire plane of $r$

\vskip 3mm
\noindent
{\it  Incoming solution on the horizon} 

We take the hypergeometric type expression for $R_{{\rm in};s}^{\n}$ 
given in Eq.(12) for the incoming solution 
on the outer horizon which is convergent except infinity. 
By the matching between the hypergeometric type solution and 
the Coulomb type solution in Eq.(23), we have the Coulomb type 
expression for $R_{{\rm in};s}^{\n}$ 
\bea
R_{{\rm in};s}^{\n}=A_se^{i\e\k}[K_{\n}(s){R}_{C;s}^{\n}+
K_{-\n-1}(s){R}_{C;s}^{-\n-1}]\;,
\ena
which is convergent in the region  $r>r_+$ . 
The solution which is defined in this way is convergent in the 
entire region of $r$ for all finite $\e$. That is, 
we use  the hypergeometric type expression in Eq.(12) around the outer horizon 
and the Coulomb type expression in Eq.(25) around infinity. 
In the intermediate 
region, we use either the expression by using ${ R}_{0;s}^{\n}$ in Eq.(15) 
or the Coulomb type expression depending on the situations. 

\vskip 3mm
\noindent
{\it Outgoing solution on the horizon}

By using $R_{{\rm in};-s}^{\n}$, 
the outgoing  solution on the outer horizon is given by 
$R_{{\rm out};-s}^{\n}\equiv \Delta^{-s}(R_{{\rm in};-s}^{\n})^*$.  
Thus,  this solution is valid in the entire region of $r$ for 
all finite $\e$. The relative normalization between solutions with 
spin weight $s$ and $-s$  is fixed automatically following the normalization 
of $R_{{\rm in};s}^{\n}$.

\vskip 3mm
\noindent
{\it Upgoing solution}

The upgoing solution is the one which satisfies the 
outgoing boundary condition at infinity so that we can 
write 
\bea
R_{{\rm up};s}^{\n}=B_sR_{-;s}^{\n}\;,
\ena 
where $B_s$'s are normalization constants and 
the  relative normalization between $B_s$ and $B_{-s}$ is fixed 
for $s>0$  with the choice $B_{-s}=1$ 
\bea
B_{-s}=1,\hskip 5mm B_{s}=C_s^*\omega^{2s}\;,   
\ena
where $C_s$'s are Starobinsky constants (defined in Eq.(49)-(53)). 
By using the relations 
\bea
R_{+;s}^{-\n-1}&=& -ie^{-i\pi\n}\frac{\sin \pi(\n-s+i\e)}{\sin \pi(\n+s-i\e)}
R_{+;s}^{\n}\;,\nonumber\\
R_{-;s}^{-\n-1}&=& ie^{i\pi\n}
R_{-;s}^{\n}\;,
\ena
we find 
\bea
R_{{\rm up};s}^{\n}=B_s\frac{e^{-\pi\e}}{\sin 2\pi\n}
\left(-i\frac{\sin\pi(\n+s-i\e)}{K_{-\n-1}(s)}R_{0;s}^{-\n-1}+
\frac{e^{-i\pi\n}\sin\pi(\n-s+i\e)}{K_\n(s)}R_{0;s}^{\n}\right)\;.
\ena
Then, if needed, we can express $R_{0;s}^{\n}$ in terms of 
$R_{{\rm in};s}^{\n}$ and $R_{{\rm out};s}^{\n}$ and then we obtain 
the hypergeometric type expression. 
Now, we obtained the upgoing solution which is written by the 
hypergeometric type expression which is convergent in the region 
$r <\infty$ and the Coulomb type expression which is convergent 
for $r>r_+$. Thus, we find the upgoing solution which is 
convergent in the entire region of $r$ for all finite $\e$.

\section{Some properties of coefficients and $K_{\n}(s)$}

In this section, we present various properties of coefficients and 
some relations between $K_{\n}(s)$'s which become important 
in the following discussion. Hereafter, we consider 
$s=0$ (masless scalar), $s=\pm 1/2$ (massless fermion, neutrino), 
$s=\pm 1$ (photon) and $s=\pm 3/2$ (massless 
spin 3/2 particle) and $s=\pm 2$ (graviton). 

\noindent
(a) Properties of coefficients

In below, we take  initial values of  coefficients 
for $s=0, \pm1/2, \pm1, \pm3/2, \pm2$   
\bea
a_0^{\n}(s)=a_0^{-\n-1}(s)=1 \;.
\ena 
Then,  from the three term recurrence relation, we have 
\bea
 a_{-n}^{-\n-1}(s)=a_n^{\n}(s)\;.
\ena
In addition, by using the explicit forms of 
$\a_n^{\n}(s)$, $\b_n^{\n}(s)$ and $\c_n^{\n}(s)$ and the three term 
recurrence relation, we obtain 
the following relations:
\bea
a_n^{\n}(-s)= \left | \frac{(\n+1+s+i\e)_n}{ (\n+1-s+i\e)_n} \right |^2
      a_n^{\n}(s),
\ena
\bea
a_n^{\n}(s)^*=(-1)^n    \frac{(\n+1+i\t)_n}{ (\n+1-i\t)_n}
      a_n^{\n}(s),
\ena
\bea
a_n^{\n}(-s)^*&=& (-1)^n \frac{(\n+1+i\t)_n}{ (\n+1-i\t)_n}
        \left | \frac{(\n+1+s+i\e)_n}{ (\n+1-s+i\e)_n} \right |^2
      a_n^{\n}(s)\;.\nonumber\\
\ena
To prove Eq.(32), we see that 
$\a_n^{\n}(-s)=|(n+\n+1-s+i\e)/(n+\n+1+s+i\e)|^2\a_n^{\n}(s)$, 
$\b_n^{\n}(-s)=\b_n^{\n}(s)$ and 
$\c_n^{\n}(-s)=|(n+\n+s+i\e)/(n+\n-s+i\e)|^2\c_n^{\n}(s)$  and 
then, it is clear that $a_n^{\n}(-s)$ defined in Eq.(32) satisfies 
the three term recurrence relation for $-s$. Similar considerations 
will lead to Eqs.(33) and (34).

 \vskip 3mm
\noindent
(b) Useful relations for $K_{\n}(s)$

The first one is 
\bea
\frac{ K_{\n}(-s)}{ K_{\n}(s)}&=&
\frac{ K_{-\n-1}(-s)}{ K_{-\n-1}(s)}\nonumber\\
&=&
\frac{1}{(\e\k)^{2s}}\frac{\G(1+s-2i\e_+)}{\G(1-s-2i\e_+)}
\left | \frac{\G(\n+1-s+i\e)}{\G(\n+1+s+i\e)} \right |^{2s}\;,
\nonumber\\
\ena
which can be proved directly by using the relation between 
$a_n^{\n}(s)$ and $a_n^{\n}(-s)$ in Eq.(32). 

The second one is 
rather nontrivial, but can be proved directly for $\tilde r=0$, 
\bea
{ K}_{\n}(s){ K}_{-\n-1}(-s)^*
& =&
\frac{2\e\k\G(1+s+2i\e_+)\G(1-s-2i\e_+)\sin \pi(\n-i\t)}{\pi} 
\nonumber\\
&&\hskip 3mm
\times \left| \frac{ \sin\pi(\n-s+i\e)}
{ \sin 2\pi \n}\right|^2\;,
\nonumber\\
\ena
where we used Eqs.(31) and (32). 
Since $K_{\n}(s)$ is independent on the choice of $\tilde r$, 
the relation in Eq.(36) is valid for all integer value of 
$\tilde r$.  
These are  useful relations which can be used to prove that 
the solution satisfies the T-S identity directly and 
to discuss  the absorption rate and the evaporation rate of black 
hole.

\section{Asymptotic behaviors }

We consider asymptotic behaviors of our solution  
on the outer horizon and at infinity. Then, we write 
asymptotic amplitudes by using our solution explicitly 
following the definition given in the book of Chandrasekhar[10]. 
The general properties derived by using the  asymptotic behaviors before  
 are compared to the explicit forms derived here. By doing 
this comparison, we can examine the structure of the solution and 
get a deep incite of our solution. The result presented here 
will be useful to derive various quantities by our solution. 

\vskip 3mm
\noindent
(a) Definition of asymptotic amplitudes 
$R_{s}^{(\rm inc)}$, $R_{s}^{(\rm ref)}$ and $R_{s}^{(\rm trans)}$

The asymptotic  behaviors are defined by[10]
\bea
R_{s}&\to& R_{s}^{(\rm inc)}\frac 1{r}e^{-i\omega r_*}+
R_{s}^{(\rm ref)}\frac 1{r^{1+2s}}e^{i\omega r_*} \hskip 5mm (r\to \infty),
\nonumber\\
 &\to& R_{s}^{(\rm trans)}\Delta^{-s}e^{-ik r_*} 
\hskip 5mm (r\to r_+),
\ena
where $k$ is defined in Eq.(4) and $r_*$ is defined by
$dr_*/dr=(r^2+a^2)/\Delta$. Then, we find 
\bea
\omega r_* &\to&  z +\e\ln  z \hskip 5mm (r \to \infty),
\nonumber\\
kr_*&\to& \e_+\ln(-x) \hskip 5mm (r \to r_+).
\nonumber\\
\ena

\noindent
(b) Asymptotic amplitudes expressed by using our solution 
$R_{{\rm in};s}^{\n}$
 
Here we express $R_{s}^{(\rm inc)}$, $R_{s}^{(\rm ref)}$ and 
$R_{s}^{(\rm trans)}$ explicitly by using the analytic  solution. 
In below, we consider the cases of  $s=0, \pm  1/2$ (neutrino), $s=\pm 1$ 
(photon), $\pm 3/2$ (massless spin 3/2 particle) 
and $\pm 2$ (graviton).

The asymptotic amplitude  
on the outer horizon $R_{s}^{(\rm trans)}$ is  
given from Eq.(12) by 
\bea
R_{s}^{(\rm trans)}&=&A_{s}\left(\frac{\e\k}{\omega}\right)^{2s}
 \sum_{n=-\infty}^{\infty}a_{n}^{\n}(s)\;.
\ena

To derive asymptotic amplitudes at infinity $R_{s}^{(\rm inc)}$ and  
$R_{s}^{(\rm ref)}$, we need some nontrivial works. We define
\bea
{ R}_{{\rm C},s}^{\n} \to A_{+;s}^{\n}{ z}^{-1}
e^{-i( z+\e\ln  z)} +
A_{-;s}^{\n}{ z}^{-1-2s}
e^{i( z+\e\ln  z)}. 
\ena 
Then, from Eq.(17), we find
\bea
A_{+;s}^{\n}&=&2^{-1+s-i\e}e^{i(\pi/2)(\n+1-s)}e^{-\pi \e/2}
\frac{\G(\n+1-s+i\e)}{\G(\n+1+s-i\e)}\sum_{n=-\infty}^{\infty} 
a_n^{\n}(s)\;,\nonumber\\
A_{-;s}^{\n}&=&2^{-1-s+i\e}e^{-i(\pi/2)(\n+1+s)}e^{-\pi \e/2}
\sum_{n=-\infty}^{\infty}(-1)^n
\frac{(\n+1+s-i\e)_n}{(\n+1-s+i\e)_n}a_n^{\n}(s)\;.\nonumber\\
\ena
Since the asymptotic behaviors of ${ R}_{{\rm in},s}^{\n}$ are 
expressed by the combination of ${ R}_{{\rm C},s}^{\n}$ and 
${ R}_{{\rm C},s}^{-\n-1}$, the asymptotic  amplitudes look 
quite involved. It is quite important that these asymptotic 
amplitudes are expressed in simpler forms 
to examine the T-S identities by using the asymptotic behaviors, 
to derive the absorption rate and the numerical computations. 
We found various useful relations. 
Firstly, we  relate $A_{+;s}^{\n}$ to $A_{+;s}^{-\n-1}$. 
By using Eq.(31), we find 
\bea
A_{+;s}^{-\n-1}&=&-ie^{-i\pi\n}\frac{\sin \pi(\n-s+i\e)}{\sin \pi(\n+s-i\e)}
A_{+;s}^{\n}\;,
\nonumber\\
A_{-;s}^{-\n-1}&=&ie^{i\pi\n}
A_{-;s}^{\n}\;.
\nonumber\\
\ena
By combining Eqs.(25), (40) and (42), we find
\bea
R_{s}^{\rm (inc)}&=&\frac{A_s e^{i\e\k}}{\omega}\left[{ K}_{\n}(s)-
 ie^{-i\pi\n} \frac{\sin \pi(\n-s+i\e)}{\sin \pi(\n+s-i\e)}
{ K}_{-\n-1}(s) \right]A_{+;s}^{\n}\;, \nonumber\\
R_{s}^{\rm (ref)}&=&\frac{A_s e^{i\e\k}}{\omega^{1+2s}}\left[{ K}_{\n}(s)+
 ie^{i\pi\n} 
{ K}_{-\n-1}(s) \right]A_{-;s}^{\n}\;. \nonumber\\
 \ena
These are expressed in compact forms. In addition, we find 
the relation between $A_{\pm;s}^{\n}$ to $A_{\pm;s}^{\n}$
by using Eq.(32) 
\bea
\frac{A_{+;s}^{\n}}{A_{+;-s}^{\n}}&=&2^{2s}e^{-i\pi s}\left|
 \frac{\G(\n+1-s+i\e)}{\G(\n+1+s+i\e)}
\right|^2 \frac{\sum_{n=-\infty}^{\infty}a_n^{\n}(s)}
{\sum_{n=-\infty}^{\infty}a_n^{\n}(-s)}
,\nonumber\\
\frac{A_{-;s}^{\n}}{A_{-;-s}^{\n}}&=&2^{-2s}e^{-i\pi s}\;.\nonumber\\
\ena
These relations are useful to discuss the T-S identities. 

\vskip 3mm
\noindent
\section{The Teukolsky-Starobinsky identities}

The Teukolsky-Starobinsky (T-S) identities[6],[7] 
for $s=1/2$, 1, 3/2 and 2 can 
be expressed as  
\bea
\Delta^s (D^{\dagger})^{2s}\Delta^s R_{s}=C_s^*R_{-s}\;,
\hskip 1cm ({\rm T-S \hskip 2mm identity \;(A)})
\ena
\bea
 (D)^{2s}R_{-s}=C_sR_{s}\;,\hskip 1cm ({\rm T-S \hskip 2mm identity \;(B)})
\ena
where $D$ and $D^{\dagger}$ are differential operators
\bea
D&=&\frac{\partial }{\partial r}-i\frac{K}{\Delta}\nonumber\\
  &=&-\frac{\omega}{\e \kappa}\left(\frac {d}{dx}+i\e\k-i\frac{\e_+}{x}
   +i\frac{\e_-}{1-x}   \right), \nonumber\\
 &=&\omega \left(\frac {d}{d z}-i-i\frac{\e_+}{ z -\e\k}
   -i\frac{\e_-}{ z}   \right),\\ 
D^{\dagger}&=&\frac{\partial }{\partial r}+i\frac{K}{\Delta}\;,
\ena
where $K=(r^2+a^2)\omega-am$. Note that $K$ used here 
corresponds to $-K$ defined in the book of Chandrasekhar[6]. 
In the above expression, 
 $(D^{\dagger})^{2s}$ means that the differential operation $D^{\dagger}$ 
applies $2s$ times and the same for $(D)^{2s}$. 

The Starobinsky constants[7]  
$C_s$' are  given as follows:  
\bea
\mid C_2\mid^2&=&(Q_2^2+4a\omega m-4a^2\omega^2)
[(Q_2-2)^2+36a\omega m-36a^2\omega^2]\nonumber\\
    &&+(2Q_2-1)(96a^2\omega^2-48a\omega m)+144\omega^2(M^2-a^2)\;,
\ena
 \bea
( C_{\frac 32})^2 &=&(Q_{\frac 32}-\frac 34)^2(Q_{\frac 32}+\frac14)
-16a^2\omega^2(Q_{\frac 32}-\frac 74)+16am\omega (
Q_{\frac 32}-\frac 34)\;,
\ena
\bea
(C_1)^2=Q_1^2-4a^2\omega^2+4am\omega\;, 
\ena
\bea
(C_{\frac 12})^2=Q_{\frac 12}+\frac14\;,
\ena
where 
\bea
Q_s=E(s)+a^2\omega^2-2a\omega m\;.\nonumber 
\ena
Trivially, $C_0=1$. 
The coefficient $C_2$ is a complex number, but others can be 
real. The coefficients $C_2$ and $C_1$ are first obtained by Teukolsky and 
Press[7] and $C_{\frac 32}$ is given by Torres del Castillo[11].

Solutions of Teukolsky equation should 
satisfy the T-S identities. It is noted that 
if a function satisfies the 
T-S relation (A), it also satisfies the T-S relation (B), 
and vice versa. This is shown  by  identity 
\bea
\Delta^s (D^{\dagger})^{2s}\Delta^s(D)^{2s}R_{-s}=
\mid C_s \mid^2 R_{-s}\;, 
\ena 
for $s =0$, 1/2, 1,  3/2, 2. The proof for $s=1$ and 2 were given 
in the book of Chandrasekhar[10] and the other cases can be proved 
similarly. 
 This fact is important for our 
solution because it may be hard to prove analytically that 
our solution satisfies both of the T-S identities, although it is 
a solution of the Teukolsky equation. This is because our solution 
is expressed as series. 

Fortunately, we can show analytically that 
$R_{{\rm in};\pm s}^{\n}$'s for $s =0$, 1/2, 1, 
 3/2, 2  satisfy the T-S identity (A). By using this, we 
can fix the relative normalization between solutions with spin weight $s$ 
 and $-s$ analytically. 
 
Firstly, we consider the hypergeometric type expression  for 
$R_{{\rm in};\pm s}^{\n}$ in Eq.(12). As we see in Appendix A, we find 
\bea
\Delta^s (D^{\dagger})^{2s}\Delta^s 
R_{{\rm in};s}^{\n}
&=&\frac{A_{s}}{A_{-s}}
\left(\frac{\e\k}{\omega}\right)^{2s}\frac{\G(1-s-2i\e_+)}{\G(1+s-2i\e_+)}
\left | \frac{\G(\n+1+s+i\e)}{\G(\n+1-s+i\e)}\right|^2 
R_{{\rm in};-s}^{\n}\;.\nonumber\\
\ena
Then, it is clear that by using the normalization factors in Eq.(13), 
the righthand side reduces to $C_s^*R_{{\rm in};-s}^{\n}$. That is, 
by taking the relative normalization in Eq.(13), 
the hypergeometric type expression for $R_{{\rm in};\pm s}^{\n}$ 
satisfies the T-S identity (A).  

In addition, we can also show that the Coulomb type expression for
$R_{{\rm in};\pm s}^{\n}$ 
satisfies the T-S relation (A) as well. 
To prove this, we first observe that  the following relation 
is satisfied:  
\bea
 \Delta^s (D^{\dagger})^{2s}\Delta^s R_{\pm;s}^{\n}
=\frac 1{\omega^{2s}}R_{\pm;-s}^{\n}\;,
\ena 
for  $s=0$, 1/2, 1, 3/2, 2.  
The proof of this relation is given  in Appendix A. 
Then, we find  from Eq.(20)
\bea
\Delta^s (D^{\dagger})^{2s}\Delta^s{R}_{C;s}^{\n}=
\frac 1{\omega^{2s}}{R}_{C;-s}^{\n}\;.
\ena
By using the Coulomb type expression in Eq.(25),  we can 
immediately find 
\bea
 \Delta^s( D^{\dagger})^{2s}\Delta^s 
R_{{\rm in};s}^{\n}
&=&\frac{A_{s} e^{i\e\k}}{\omega^{2s}}
\left(K_{\n}(s)R_{C;-s}^{\n}+K_{-\n-1}(s)R_{C;-s}^{-\n-1}\right)
\nonumber\\
&=&C_s^*R_{{\rm in};-s}^{\n}\;,
\ena
where we used the relation among $K_{\n}(s)$ given in Eq.(35) and the 
relative normalization in Eq.(13) to derive the last equality. 
In summary, both expressions of $R_{{\rm in};s}^{\n}$  
satisfies the T-S identity (A). 

It is not possible to prove analytically that our solution  
$R_{{\rm in};s}^{\n}$  
satisfies the T-S identity (B), although it should satisfy by the 
identity in Eq.(53).  It is expected that  the T-S 
identity (B) will give some nontrivial identities if it  applies to 
our solution. 

The analysis by using  the asymptotic behaviors of the solution 
has been made by Teukolsky and Press[7]. By substituting 
the asymptotic behaviors in Eq.(37) on the outer horizon 
into the the  T-S identity (B), one finds   for 
$s =0$, 1/2, 1,  3/2, 2 
\bea
R_{s}^{({\rm trans})}&=&\frac 1{C_s}\left(\frac{\e\k}{\omega}\right)^{2s}
\frac{\G(1+s-2i\e_+)}{\G(1-s-2i\e_+)}
R_{-s}^{({\rm trans})}\;.
\ena
Similarly, by considering the T-S identity (A) and (B) at infinity, 
one finds
\bea
R_{s}^{({\rm inc})}&=&e^{-i\pi s}\frac {(2 \omega)^{2s}}
{C_s}R_{-s}^{({\rm inc})}\;,\nonumber\\
R_{s}^{({\rm ref})}&=&e^{-i\pi s}\frac {C_s^*}{(2\omega)^{2s}}
R_{-s}^{({\rm ref})}\;,\nonumber\\
\ena
The relations in Eq.(58) and (59) are essentially the same as those 
by Teukolsky and Press[7]. By requiring that 
the asymptotic amplitudes in Eqs.(39) and (43) satisfy 
the Teukolsky-Press relations, we can derive some identities 
between the sums of coefficients which we shall see in Sec.7. 

Finally by using the relation in Eq.(55), we can show that the upgoing 
solution with the relative normalization in Eq.(27) 
satisfies the T-S identity (A). That is, we find  
\bea
\Delta^s (D^{\dagger})^{2s}\Delta^sR_{{\rm up};s}^{\n}
&=&B_s\Delta^s (D^{\dagger})^{2s}\Delta^sR_{-;s}^{\n}
\nonumber\\
&=&C_s^*R_{{\rm up};-s}^{\n}\;. 
\ena 
Thus,  the relative normalization between the solution with spin weight 
$s$ and 
$-s$ is determined analytically for the upgoing solution as given in Eq.(27). 

\section{The conserved current and the energy conservation}

In below, we take $s$ to be  0, 1/2, 1, 3/2, 2. 
The  Wronskian gives the conserved current[7] which  is 
written by 
\bea
\left[ Y(-s)^*\frac{dY(s)}{dr_*}-Y(s)\frac{dY(-s)^*}{dr_*}\right ]_{r=r_+}
&=&\left[ Y(-s)^*\frac{dY(s)}{dr_*}-Y(s)\frac{dY(-s)^*}{dr_*}
\right ]_{r=\infty}\;,\nonumber\\
\ena 
where  $Y(s)=\Delta^{s/2}(r^2+a^2)^{1/2}R_{{\rm in};s}$. 
By substituting the asymptotic behaviors in Eq.(37), one finds 
\bea
(R_{s}^{\rm (inc)})(R_{-s}^{\rm(inc)})^*=
(R_{s}^{\rm(ref)})(R_{-s}^{\rm(ref)})^* 
-i\frac{\e\k (s+2i\e_+)}{2\omega^2}(R_{s}^{\rm(trans)})
(R_{-s}^{\rm(trans)})^*\;.
\ena
By using the Teukolsky-Press relations 
in Eqs.(58) and (59), we obtain 
\bea
|R_{s}^{\rm(inc)}|^2=
\frac{(2\omega)^{4s}}{|C_s|^2}|R_{s}^{\rm(ref)}|^2 
+\delta_s |R_{s}^{\rm(trans)}|^2,
\ena
where 
\bea
\delta_s=-ie^{i\pi s}\omega^{4s-2}\left(\frac{\e\k}{2}\right)^{-2s+1}
\frac{\G(1-s+2i\e_+)}{\G(s+2i\e_+)}\;.
\ena
Explicitly, we obtain
\bea
\delta_0&=&\frac{\e_+(\e\k)}{\omega^2}\;,
\nonumber\\
\delta_{\frac 12}&=&1\;.
\nonumber\\
\delta_1&=&\frac{\omega^2}{\e_+(\e\k)}\;,
\nonumber\\
\delta_{\frac 32}&=&\frac{4\omega^4}{(\frac14+4\e_+^2)(\e\k)^2}\;,
\nonumber\\
\delta_2&=&\frac{4\omega^6}{\e_+(1+4\e_+^2)(\e\k)^3}\;,
\nonumber\\
\ena
The  relation in Eq.(63) leads to  the energy conservation[6],[10] 
\bea
\frac{d^2E_{\rm inc;s}}{dtd\Omega}=\frac{d^2E_{\rm ref;s}}{dt
d\Omega}+\frac{d^2E_{\rm trans;s}}{dtd\Omega}\;,
\ena 
where $d^2E_{\rm inc;s}/{dtd\Omega}$ is the incident energy 
going into the black hole, $d^2E_{\rm ref;s}/{dtd\Omega}$ is the 
energy reflected by the black hole and 
$d^2E_{\rm trans;s}/dtd\Omega$ is the energy absorbed by the 
black hole. 
It is clear that $\delta_s$'s are proportional to  
$\e_+$ for bosons and are positive definite for fermions. Since 
$2\e_+=\e+\t=\e(1+\frac 1{\k})-\frac{ma}{M\k}$, $\e_+$ 
can be negative for large angular momentum of black 
hole. That is,
the super radiance occurs for bosons for $a>\frac{\e(1+\k)M}{m})$, but not 
fermions[12].

By substituting the explicit forms of  asymptotic amplitudes 
in Eqs.(39) and  (43), we have an identity involving the 
sums of coefficients which we shall see in the next section. 
In Sec.8, we shall derive a simple expression for the 
absorption rate.

\section{Identities involving the sums of coefficients} 

The requirement that the asymptotic amplitude on the outer horizon 
 in Eq.(39) satisfies the Teukolsky-Press relation in Eq.(58)   
leads to  the  identity  for $s=0$, 1/2, 1, 3/2, 2
\bea
\sum_{n=-\infty}^{\infty} a_n^{\n}(-s)
=
|C_s|^2 \left |\frac{\G(\n+1-s+i\e)}{\G(\n+1+s+i\e)}  \right |^2
{\sum_{n=-\infty}^{\infty} a_n^{\n}(s)}
 \;. \hskip 1cm
(I-1)\nonumber\\
\ena
The requirement that the asymptotic amplitudes at infinity in 
Eq.(43) satisfy the Teukolsky-Press relations in Eq.(59) is 
automatically satisfied. This can be seen by using the relation 
among $K_{\n}(s)$ in Eq.(35). No more identity arises from 
the T-S identities. 

Next, we consider the constraint from the Wronskian. 
By substituting the asymptotic amplitudes in Eqs.(39) and (43) 
into Eq.(63) and by using the identity (I-1), we find 
\bea
\frac{\left|\sum_{n=-\infty}^{\infty}(-1)^n\frac{(\n+1+s-i\e)_n}{
(\n+1-s+i\e)_n} a_n^{\n}(s)\right|^2}{
\left|\sum_{n=-\infty}^{\infty} a_n^{\n}(s)\right|^2}=
|C_s|^2 \left |\frac{\G(\n+1-s+i\e)}{\G(\n+1+s+i\e)}  \right |^2
 \;. \hskip 0.5cm
(I-2)\nonumber\\
\ena
Since the derivation is lengthy, we shall give a brief proof in 
Appendix A. The identity (I-2) can be written in a different form 
by using the identity (I-1) as 
\bea
\left|\sum_{n=-\infty}^{\infty}(-1)^n\frac{(\n+1+s-i\e)_n}{
(\n+1-s+i\e)_n} a_n^{\n}(s)\right|^2=
\left(\sum_{n=-\infty}^{\infty} a_n^{\n}(s)\right)
\left(\sum_{n=-\infty}^{\infty} a_n^{\n}(-s)\right)^*\;. \hskip 0.5cm
(I-2)'\nonumber\\
\ena

The identities (I-1) and (I-2) have simple structures 
and will serve to make various formulas into simpler ones. 
Since our solution is given by the sums of either hypergeometric or 
Coulomb wave functions, various formulas given by our solution are 
expressed in complicated  forms involving  the sums of coefficients. 
Therefore, these identities are powerful to compute various 
quantities analytically and also numerically. 

Another important feature of  these identities lies in the fact that 
they can be used as a good tool  
to test the accuracy of the computation. In order to obtain the Teukolsky 
function, we have to solve the three term recurrence relation by 
the $\e$ expansion or the numerical calculation. In the computation, 
the estimation of the accuracy becomes an important issue. 
The accuracy  is directly 
tested by looking how accurately these identities are satisfied 
because the sums appearing in these identities correspond to  the asymptotic 
amplitudes themselves. 
As for the accuracy test of the computation,  
to check the $\tilde r$ independence of $K_{\n}(s)$ will 
be another good test.  
These identities and relation give nontrivial test of the accuracy of 
computation. 
  
\section{The absorption rate and the evaporation rate of black hole}

In this section, we present  the  absorption rate formula expressed 
by using our solution. Since the calculation is lengthty, we give 
the brief derivation in Appendix A and only give the result 
\bea
\G_s&=&\delta_s \left|\frac{R_{s}^{\rm(trans)}}{R_{s}^{\rm(inc)}}\right|^2
=1-\frac{(2\omega)^{4s}}{|C_s|^2}
\left|\frac{R_{s}^{\rm(ref)}}{R_{s}^{\rm(inc)}}\right|^2
\nonumber\\
&=&(2\e\k)^{2\n+1}\frac{e^{\pi\e}}{\pi}\left\{
\matrix{\sinh 2\pi\e_+\cr \cosh 2\pi \e_+\cr}\right\}
\nonumber\\
&& \times
\frac{D_s^{\n}}{
\left|1+i\frac{(2\e\k)^{2\n+1}e^{i\pi\n}(-1)^{2s}\sin\pi(\n-i\t)}{\pi}
\left(\frac{\sin\pi(\n-s-i\e)}
{\sin2\pi\n}\right)^2 D_s^{\n}\right|^2}\;,
\nonumber\\
\ena
where the upper column ($\sinh 2\pi \e_+$) is taken for bosons ($s=0,1,2$) 
 and the lower column ($\cosh 2\pi\e_+$) is taken 
for fermions ($s=1/2,3/2$). Here,   
\bea
D_s^{\n}=\left|\frac{\G(\n+1-i\t)\G(\n+1-s+i\e)\G(\n+1+s+i\e)}
{\G(2\n+1)\G(2\n+2)}\right|^2 d_s^{\n}
\nonumber\\
\ena
where
\bea
d_s^{\n}&=&\left|\sum_{n=-\infty}^{0}\frac{(-1)^n}{(-n)!(2\n+2)_n}
\frac{(\n+1+s-i\e)_n}{(\n+1-s+i\e)_n}a_n^{\n}(s)\right|^2
\nonumber\\
&&\hskip 3mm \times
\left| \sum_{n=0}^{\infty}\frac{(2\n+1)_n}{n!}
\frac{(\n+1+s-i\e)_n}{(\n+1-s+i\e)_n}a_n^{\n}(s)\right|^{-2}\;.
\ena
To derive the above formula, we used 
$\sin\pi(\n+s-i\e)=(-1)^{2s}\sin\pi(\n-s-i\e)$. 
It may be worthwhile to note that   
$d_s^{\n}$ starts from  1 in the  $\e$ expansion and is the only part which 
we have make a calculation by solving the three term recurrence relation. 
Since the mathematical structure is simple, it is easy to estimate 
the rate either in the expansion of $\e$ or the direct numerical 
calculation.

The evaporation rate from the 
black hole is given by[13] 
\bea
<N>&\equiv& \frac{\G_s}{e^{\frac{2\pi}{\k'}(\omega-m\Omega_+)}\mp 1}
\nonumber\\
&=&
2\left\{\matrix{(\sinh 2\pi\e_+)^{-1}\cr (\cosh 2\pi\e_+)^{-1}\cr} 
   \right\} e^{-\frac{\pi}{\k'}(\omega-m\Omega_+)}\G_s\;.
\ena 
Here, the upper column for bosons and the lower column for fermions, 
$\k'$ is surface gravity, $\Omega_+$ is angular velocity on 
horizon and we used 
\bea
2\e_+&=&\e+\t\equiv 2M\omega +\frac{2M\omega-m\frac{a}{M}}
{\sqrt{1-\left(\frac{a}{M}\right)^2}}
\nonumber\\
&=&\frac{\omega-m\Omega_+}{\k'}\;.
\ena
By substituting $\G_s$ in Eq.(70) into the  evaporation rate formula,  
we find 
\bea
<N>&=&\frac 2{\pi} e^{-\frac{\pi}{\k'}(\frac{\omega}{1+\k}-m\Omega_+)}
(2\e\k)^{2\n+1}
\nonumber\\
&& \times
\frac{D_s^{\n}}{
\left|1+i\frac{(2\e\k)^{2\n+1}e^{i\pi\n}(-1)^{2s}\sin\pi(\n-i\t)}{\pi}
\left(\frac{\sin\pi(\n-s-i\e)}
{\sin2\pi\n}\right)^2 D_s^{\n}\right|^2}\;.
\ena
It is interesting to observe that 
the denominator in the evaporation $<N>$ which shows the difference between 
the boson and the fermion emissions is canceled by the $\sinh 2\e_+$ or 
$\cosh 2\e_+$ factor in $\G_s$. As a result, 
the evaporation rate  shows 
the Boltzman like behavior both for both boson and fermion emissions.  

In below, we made the $\e$ expansion for $\G_s$. 
In Ref.1, we obtained the renormalized angular 
momentum $\n$ and coefficients up to the order $\e^2$. 
In the approximation neglecting the   $O(\e^2)$ terms, 
we can immediately derive the absorption rate because  
$\n=l+O(\e^2)$. We find 
\bea
\G_s&=& (2\e\k)^{2l+1}e^{\pi\e}
\left\{
\matrix{\frac{\t\sinh \pi(\e+\t)}{\sinh \pi\t}\Pi_{k=1}^{l}(k^2
+\t^2)
\cr 
\frac{\cosh \pi( \e+\t)}{\cosh\pi\t}
\Pi_{k=1/2}^{l}(k^2
+\t^2)
\cr}\right\}
\nonumber\\
&&\hskip 5mm \times
\left(\frac{(l-s)!(l+s)!}{2l!(2l+1)!}\right)^2 d_s^{\n}\;,
\ena
where the upper cases for massless scalar, photons and 
gravitons, the lower case 
for neutrinos and massless spin 3/2 particles and 
\bea
d_s^{\n}=1+ mqs^2\left(\frac 1{l^2}-\frac 1{(l+1)^2}\right)\e \;.
\ena
For massless scalar case ($s=0$), it is understood that 
$\Pi_{k=1}^{l}(k^2+\t^2)=1$ for $l=0$. In the zeroth order of $\e$, it is 
easy to see that this expression reproduces 
the formulas first given by  Page[14] for $s=0$, 1/2, 1 and 2.

\section{Summary}

To obtain the Teukolsky functions in a desired accuracy is a quite 
important issue for the gravitational wave astrophysics. The analytic 
solution which we reported in our previous paper is expected to 
serve for this purpose. Since the solution is given as series of 
special functions, the examination of the property of the solution 
is needed. 

In this paper, we used new Coulomb type solutions and constructed 
the analytic solutions valid in the entire region of $r$ for all 
finite $\e$. By using this solution, we  examined  various properties 
of solutions of 
Teukolsky equation. We found that the relative normalization 
between the solution with a spin weight $s$ and $-s$ is determined 
analytically by using the T-S identities. In addition, we found 
an identity (I-1) involving the sums of coefficients of series. 
By using the Wronskian, we found another identity (I-2). The sums 
of coefficients appeared in these identities are proportional to 
the asymptotic amplitudes on the horizon and at infinity and 
thus can be used to test the accuracy for the computation. The 
remarkable fact is that we can compute the Teukolsky functions 
systematically in a desired accuracy. By using these identities, 
we demonstrated that the asymptotic amplitudes can be expressed 
in compact forms. The fact that various quantities can be written 
simpler forms by using these identities is important to see the 
general properties of these quantities analytically without any 
approximation and also to make the numerical computation. 

As an example, we derived the absorption rate and the evaporation 
rate of black hole. We found that they are expressed in compact 
forms from which we can extract the physics transparently. 

Now, the properties of solutions are well understood and it is ready 
to go to derive the Teukolsky functions by either $\e$ expansion or the 
numerical computation which we are now pursuing.  

\vskip 1cm
{\Huge Acknowledgment}

We would like to thank to H. Suzuki for discussions, M. Sasaki, 
M. Shibata, H. Tagoshi and T. Tanaka for comments and 
encouragements. This work is supported in part by 
the Japanese Grant-in-Aid for Scientific Research of
Ministry of Education, Science, Sports and Culture, 
No. 06640396 and 08640374.
  
\newpage

%Appendix format setting
\setcounter{section}{0}
\renewcommand{\thesection}{\Alph{section}}
\renewcommand{\theequation}{\thesection .\arabic{equation}}
\newcommand{\apsc}[1]{\stepcounter{section}\noindent
\setcounter{equation}{0}{\Large{\bf{Appendix\,\thesection:\,{#1}}}}}

\apsc{Proofs of various formulas}

In this appendix, we give brief proofs of various formulas 
given in the text. 
\vskip 3mm
\noindent
(a) The Teukolsky-Starobinsky identities

\noindent
{\it Proof of equation (54)}

For $s=0$, 1/2, 1, 3/2, 2, we rewrite $R_{{\rm in};s}^{\n}$ as 
\bea
R_{{\rm in};s}^{\n}&=&A_{s}e^{i\e\kappa x}(-x)^{-s-i\e_+}
(1-x)^{-s-i\e_-}
\left\{
(1-x)^{s+2i\e_-}\sum_{n=-\infty}^{\infty}a_n^{\n}(s)\right. \nonumber\\
&&\hskip 3mm \left.\times F(n+\n+1-i\t,-n-\n-i\t;1-s-2i\e_+;x)\right \}\;.
\nonumber\\
\ena
Then, one finds 
\bea
&&\Delta^s (D^{\dagger})^{2s}\Delta^s
R_{{\rm in};s}\nonumber\\
&&\hskip 0.5cm =A_{s}\left(-\frac{\e\k}{\omega}\right)^{2s}
e^{i\e\kappa x}(-x)^{s-i\e_+}(1-x)^{s-i\e_-}
\frac{d^{2s}}{dx^{2s}} \left\{
(1-x)^{s+2i\e_-}\sum_{n=-\infty}^{\infty}a_n^{\n}(s)\right. \nonumber\\
&&\hskip 6mm \times \left. F(n+\n+1-i\t,-n-\n-i\t;1-s-2i\e_+;x)
\right \}\;.
\nonumber\\
\ena
Here we use the relation among hypergeometric function 
\bea
&&\frac{d^n}{dx^n}(1-x)^{a+b-c}F(a,b;c;x)\nonumber\\
&&\hskip 1cm =\frac{(c-a)_n(c-b)_n}{c_n}
(1-x)^{a+b-c-n}F(a,b;c+n;x)
\nonumber\\
\ena
and then the right hand side of Eq.(A.3) becomes 
\bea
&&A_{s}\left(-\frac{\e\k}{\omega}\right)^{2s}
e^{i\e\kappa x}(-x)^{s-i\e_+}(1-x)^{i\e_-}\nonumber\\
&& \hskip 2cm \times \sum_{n=-\infty}^{\infty}a_n^{\n}(s)
\frac{(-n-\n-s-i\e)_{2s}(n+\n+1-s-i\e)_{2s}}{(1-s-2i\e_-)_{2s}}
\nonumber\\
&&\hskip 2cm \times
F(n+\n+1-i\t,-n-\n-i\t;1+s-2i\e_+;x)\;.
\nonumber\\
\ena
By using 
\bea
&&(-n-\n-s-i\e)_{2s}(n+\n+1-s-i\e)_{2s}\nonumber\\
&&\hskip 10mm =(-1)^{2s}
\left |\frac {\G(\n+1+s+i\e)}{\G(\n+1-s+i\e)}\right|^2
\left |\frac {(\n+1+s+i\e)_n}{(\n+1-s+i\e)_n}\right|^2\;
\nonumber\\
\ena
and the relation between $a_n^{\n}(s)$ and $a_n^{\n}(-s)$ in Eq.(32),  
Eq.(54) is proven .

\vskip 5mm
\noindent
{\it Proof of Eq.(55)}

The outline of the proof is as follows: 
\bea
&&\omega^{4s}2^{-\n}e^{\pi\e}e^{-i\pi(\n+1-s)} 
\frac{\G(\n+1-s+i\e)}{\G(\n+1+s-i\e)}
\Delta^s(D^{\dagger})^{2s}\Delta^s R_{+;s}^{\n}
\nonumber\\
&&\hskip 3mm =
z^s(z-\e\k)^s(D^{\dagger})^{2s}e^{-iz}z^{-i\e_-}(z-\e\k)^{-i\e_+}
\sum_{n=-\infty}^{\infty} i^na_n^{\n}(s)\nonumber\\
&&\hskip 5mm \times z^{\n+s+i\e}(2z)^n\Psi(n+\n+1-s+i\e,2n+2\n+2;2iz)
\nonumber\\
&&\hskip 3mm = \omega^{2s} 
e^{-iz}z^{s-i\e_-}(z-\e\k)^{s-i\e_+}
\sum_{n=-\infty}^{\infty} i^na_n^{\n}(s)\nonumber\\
&&\hskip 5mm \times
\frac{d^{2s}}{dz^{2s}}z^{\n+s+i\e}(2z)^n\Psi(n+\n+1-s+i\e,2n+2\n+2;2iz)
\nonumber\\
&& \hskip 3mm = \omega^{2s} 
e^{-iz}z^{\n+i\e_+}(z-\e\k)^{s-i\e_+}
\sum_{n=-\infty}^{\infty} i^na_n^{\n}(s)(n+\n+1-s+i\e)_{2s}
\nonumber\\
&&\hskip 5mm\times(-n-\n-s+i\e)_{2s}
(2z)^n\Psi(n+\n+1+s+i\e,2n+2\n+2;2iz)\;,
\nonumber\\
\ena
where we used the  relation of confluent hypergeometric functions 
\bea
\frac{d^n}{dx^n}\left(x^{a+n-1}\Psi(a,c;x)\right)&=&
(a)_n(a-c+1)_n x^{a-1}\Psi(a+n,c;x)
\nonumber\\
\ena
to derive the last line. 
By a careful calculation by using Eq.(A.6) and  the relation between 
$a_n^{\n}(s)$ and $a_n^{\n}(-s)$ in Eq.(32), we can derive 
the relation for $R_{+;s}^{\n}$ in Eq.(55).
\newpage
Similarly, we find
\bea
&&\omega^{2s}2^{-\n}e^{-\pi\e}e^{-i\pi(\n+1+s)} 
\Delta^s(D^{\dagger})^{2s}\Delta^s R_{-;s}^{\n}
\nonumber\\
&& \hskip 3mm = 
e^{-iz}z^{s-i\e_-}(z-\e\k)^{s-i\e_+}
\sum_{n=-\infty}^{\infty} i^na_n^{\n}(s)
\frac{(\n+1+s-i\e)_n}{(\n+1-s+i\e)_n}
\nonumber\\
&& \hskip 5mm \times
\frac{d^{2s}}{dz^{2s}}e^{2iz}z^{\n+s+i\e}(2z)^n\Psi(n+\n+1+s-i\e,2n+2\n+2;-2iz)
\nonumber\\
&& \hskip 3mm = (-1)^{2s}
e^{iz}z^{\n+i\e_+}(z-\e\k)^{s-i\e_+}
\sum_{n=-\infty}^{\infty} i^na_n^{\n}(s)\frac{(\n+1+s-i\e)_n}{(\n+1-s+i\e)_n}
\nonumber\\
&& \hskip 5mm \times
(2z)^n\Psi(n+\n+1-s-i\e,2n+2\n+2;-2iz)\;,
\nonumber\\
\ena
where we used the relation
\bea
\frac{d^n}{dx^n}\left(e^{-x}x^{c-a+n-1}\Psi(a,c;x)\right)&=&
(-1)^n e^{-x} x^{c-a-1}\Psi(a-n,c;x).
\nonumber\\
\ena 
By using the relation between $a_n^{\n}(s)$ and $a_n^{\n}(-s)$ in Eq.(32), 
we can derive the relation for $R_{-;s}^{\n}$ in Eq.(55).

\vskip 3mm
\noindent
(b) The derivation of the absorption rate

We find from Eq.(24) with $\tilde r=0$
\bea
|K_{\n}(s)|^2&=&\frac{\pi}{2^{2s}}\frac{(2\e\k)^{-2\n+2s}}
{\sin \pi(s+2i\e_+)}\frac{\G(1-s+2i\e_+)}{\G(s+2i\e_+)}
\left|\frac{\G(\n+1+s+i\e)}{\G(\n+1-s+i\e)}\right|^2
\frac 1{D_s^{\n}}\;,
\nonumber\\
\ena
where $D_s^{\n}$ is defined in Eq.(71).  
Then, by changing $\n$ to $-\n-1$ and using $a_{-n}^{-\n-1}(s)=a_n^{\n}(s)$, 
we find
\bea
\frac{|K_{-\n-1}(s)|^2}{|K_{\n}(s)|^2}=
\frac{(2\e\k)^{4\n+2}}
{\pi^2}
\left| \sin\pi(\n+i\t)\left(\frac{\sin\pi(\n+s+i\e)}{\sin 2\pi\n}\right)^2
\right|^2(D_s^{\n})^2\;.
\nonumber\\
\ena
\newpage
We also find  by using Eqs.(35) and (36)
\bea
\frac{K_{\n}(s)K_{-\n-1}(s)^*}{|K_{\n}(s)|^2}&=& 
\frac{K_{\n}(s)K_{-\n-1}(-s)^*}{|K_{\n}(s)|^2}\left(\frac{K_{-\n-1}(s)}
{K_{-\n-1}(-s)}\right)^*\nonumber\\
&=&
\frac{(2\e\k)^{2\n+1}\sin \pi(\n-i\t)}{\pi}
\left|\frac{\sin\pi(\n-s+i\e)}{\sin 2\pi\n}\right|^2 D_s^{\n}
\;.
\nonumber\\
\ena
By using Eqs.(A.11) and (A.12), we obtain
\bea
&&\left|{ K}_{\n}(s)-
 ie^{-i\pi\n} \frac{\sin \pi(\n-s+i\e)}{\sin \pi(\n+s-i\e)}
{ K}_{-\n-1}(s) \right|^2 \nonumber\\
&=&\frac1{D_s^{\n}}\frac{\pi}{2^{2s}}\frac{(2\e\k)^{-2\n+2s}}
{\sin \pi(s+2i\e_+)}\frac{\G(1-s+2i\e_+)}{\G(s+2i\e_+)}
\left|\frac{\G(\n+1+s+i\e)}{\G(\n+1-s+i\e)}\right|^2\nonumber\\
&\times& 
\left |1+
\frac{i(-1)^{2s}e^{i\pi\n}(2\e\k)^{2\n+1}\sin \pi(\n-i\t)}{\pi}
\left(\frac{\sin\pi(\n-s+i\e)}{\sin 2\pi\n}\right)^2 D_s^{\n}
\right|^2\;.
\nonumber\\
\ena
By using this formula, we can derive the formula for the absorption rate 
in Eq.(70).
 
\vskip 3mm
\noindent
(c) Identity (I-2) form the Wronskian

Here we show  that the identity (I-2) is obtained from the 
Wronskian given in Eq.(63). Let us first parameterize 
$|R_s^{\rm (inc)}|^2$  as
\bea
|R_s^{\rm (inc)}|^2&=&\frac{|A_s|^22^{2s-2}e^{-\pi\e}}{\omega^2} 
\{|K_{\n}(s)+ie^{i\pi\n}K_{-\n-1}(s)|^2
\nonumber\\
&&\hskip 2mm
+\left\{\left|K_{\n}(s)-ie^{-i\pi\n}\frac{\sin\pi(\n-s-i\e)}{\sin\pi(\n+s+i\e)}
K_{-\n-1}(s)\right|^2\right.\nonumber\\
&&\hskip 2mm \left.
-|K_{\n}(s)+ie^{i\pi\n}K_{-\n-1}(s)|^2\right\}
\left|\frac{\G(\n+1-s+i\e)}{\G(\n+1+s+i\e)}\right|^2
\left|\sum_{n=-\infty}^{\infty}a_n^{\n}(s)\right|^2\;.
\nonumber\\
\ena
Then, by using  
the  relation
\bea
&&|K_{\n}(s)-ie^{-i\pi\n}\frac{\sin\pi(\n-s-i\e)}{\sin\pi(\n+s+i\e)}
K_{-\n-1}(s)|^2-
|K_{\n}(s)+ie^{i\pi\n}
K_{-\n-1}(s)|^2
\nonumber\\
&&\hskip 1cm 
=-2ie^{i\pi s}(\e\k)^{2s+1}e^{\pi\e}
\frac{\G(1-s+2i\e_+)}{\G(s+2i\e_+)}
\left|\frac{\G(\n+1+s+i\e)}{\G(\n+1-s+i\e)}\right|^2,
\ena
we can prove that the contribution from the second part 
in the parenthesis in Eq.(A.14), i.e., the part in Eq.(A.15)  cancels   
the transition (absorption) part $\delta_s|R_{s}^{\rm(trans)}|^2$. 
Thus, the contribution from the 
first part in Eq.(A.14) should be equal the reflection part. 
By equating these two, we 
can derive the identity (I-2).

\end{document}